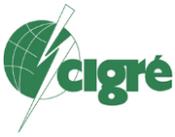

# Evaluation of Real-Time Mitigation Techniques for Cyber Security in IEC 61850 / IEC 62351 Substations

A. HERATH[1]*, C. C. LIU[1], J. HONG[2], K. PARK[2]
Virginia Polytechnic Institute and State University[1], University of Michigan-Dearborn[2]
USA

**SUMMARY**

Modern power systems are rapidly transitioning from hard-wired architectures to IEC 61850-based digital substations, where protection and control functions rely on networked communication over Ethernet. This transition improves interoperability, reduces engineering and maintenance costs, and enhances protection and control performance through fast event-driven communication and time synchronization. However, the digitalization of substations also enlarges the cyber-attack surface, necessitating effective detection and mitigation of cyber attacks in digital substations.

While machine learning-based intrusion detection has been widely explored, such methods have not demonstrated detection and mitigation within the required real-time budget. In contrast, cryptographic authentication has emerged as a practical candidate for real-time cyber defense, as specified in IEC 62351. In addition, lightweight rule-based intrusion detection that validates IEC 61850 semantics can provide specification-based detection of anomalous or malicious traffic with minimal processing delay.

This paper discusses the security vulnerabilities of IEC 61850 Generic Object-Oriented Substation Event (GOOSE) communication and provides detailed explanations of four GOOSE-based cyber attacks: replay attack, masquerade attack, flooding denial-of-service (DoS), and packet drop DoS. It further presents the design logic and implementation aspects of three potential real-time mitigation techniques capable of countering GOOSE-based attacks: (i) IEC 62351-compliant message authentication code (MAC) scheme, (ii) a semantics-enforced rule-based intrusion detection system (IDS), and (iii) a hybrid approach integrating both MAC verification and Intrusion Detection System (IDS). A comparative evaluation of these real-time mitigation approaches is conducted using a cyber-physical system (CPS) security testbed. To support this evaluation, a simulated IED is developed to replicate protection and control functions while implementing the techniques within a Mininet environment for GOOSE packet forwarding. Detection and mitigation effectiveness, as well as processing latencies, are systematically measured.

The results show that both MAC verification and semantics-enforced rule-based IDS have limitations when applied individually, whereas their hybrid integration significantly enhances mitigation capability. Furthermore, the processing delays of all three methods remain within

akilaasansana@vt.edu

the strict delivery requirements of GOOSE communication. The study also identifies limitations that none of the techniques can fully address, highlighting areas for future work.

In summary, this study demonstrates that IEC 62351 MAC verification, semantics-enforced rule-based IDS, and especially their hybrid integration are practical and effective real-time mitigation solutions against a range of GOOSE-based cyber attacks. The findings contribute to advancing the cyber security resilience of digital substations by providing experimentally validated techniques that ensure both strong security and compliance with real-time requirements.

**KEYWORDS**

Cyber Security, Digital Substation, Detection, Mitigation, IEC 61850, IEC 62351.

Acknowledgement- This research is supported by the Office of Cybersecurity, Energy Security, and Emergency Response, Cybersecurity for Energy Delivery Systems Program, of the U.S. Department of Energy, under contract DE-CR0000021.

## 1. INTRODUCTION

The modernization of power systems has accelerated the adoption of IEC 61850-based digital substations, replacing legacy, hard-wired designs and vendor-specific protocols with interoperable, Ethernet-based architectures. IEC 61850 is an international standard developed by IEC TC 57 for the automation of electrical substations, providing a comprehensive framework for communication, information modelling, and system architecture. However, critical IEC 61850 messages such as Generic Object-Oriented Substation Event (GOOSE) and Sampled Values (SV), lack built-in security and are vulnerable to cyber attacks. The strict millisecond-level delivery requirements significantly constrain feasible defense mechanisms.

A substantial amount of research has been focused on intrusion detection for IEC 61850 SV and GOOSE in the recent past. Machine-learning methods are widely explored, but they are not designed to meet the delivery time requirements for GOOSE communication and real-time mitigation. As a result, the only approaches that currently support real-time mitigation are (i) cryptographic authentication and (ii) lightweight, rule-based checks of packet semantics. IEC 62351 standard specifies cryptographic solutions that ensure data integrity and authenticity, with message authentication code (MAC)-based methods capable of meeting the sub-millisecond delivery requirements [1]. In contrast, specification-based or rule-based detection methods validate protocol semantics such as sequence numbers, state transitions, and field consistency to identify anomalies [2]. Since rule-based methods are deterministic and require minimal computation, packet forwarding based on their verification can be performed efficiently. Furthermore, combining MAC verification with specification-based detection provides fast, in-depth defense protection suitable for digital substations [3]. In fact, IEC 62351 standard recommends the integration of rule-based checks with MAC verification to improve the overall cyber attack defense coverage [4].

This paper presents a comparative evaluation of real-time mitigation approaches for GOOSE-based cyber attacks using a cyber-physical system (CPS) security testbed for substation automation. In this study, three methods are implemented and assessed: (i) an IEC 62351 compliant MAC verification, (ii) a rule-based IDS that enforces GOOSE semantics, and (iii) a hybrid approach that integrates both. Across several representative GOOSE attack types, the detection and mitigation coverage of each technique is evaluated. In addition, the latencies of the mitigation techniques are measured. From the results, each method's practical applicability in real-time cyber-attack mitigation is evaluated.

In the remainder of this paper, Section 2 presents a detailed discussion on the security vulnerabilities of GOOSE communication. A set of GOOSE-based attack scenarios is introduced in Section 3. The tested mitigation methods are defined in Section 4. Experimental results obtained from the CPS security testbed and discussions are given in Section 5, while the concluding remarks are presented in Section 6.

## 2. SECURITY VULNERABILITIES IN IEC 61850 GOOSE

GOOSE is an event-driven, publisher–subscriber messaging service used for fast protection and control in IEC 61850 digital substations. Defined at the service level in IEC 61850-7-2 and mapped directly to Ethernet in IEC 61850-8-1, GOOSE frames are Layer-2 multicast with VLAN tagging and priority, enabling fast delivery. Typical applications are illustrated in Figure 1: (1) rapid status/event publishing from process-level to protection and control IEDs (P&C IEDs), (2) command messages from P&C IEDs to process level devices, and (3) peer-to-peer status and command exchange among neighboring P&C IEDs.

In their standard form, GOOSE messages provide no built-in security. Because they operate at Ethernet Layer 2, they also lack the IP-level protection that Manufacturing Message

Specification (MMS) defined in IEC 61850 can leverage. Consequently, if an adversary gains access to the substation local area network (LAN), GOOSE traffic can be attacked across the three flow patterns described earlier. If status information from merging unit IEDs (MUIEDs) to P&C IEDs is compromised, bay-level, station-level, and control center systems may misinterpret process-level conditions and issue erroneous control actions. Attacks on command traffic from P&C IEDs to MUIEDs can directly trigger illegitimate breaker operations. Malicious peer-to-peer exchanges among IEDs can propagate misinformation and induce improper interlocking or tripping. Overall, these manipulations can destabilize the system, and in the worst case, contribute to cascading events unless additional safeguards are deployed.

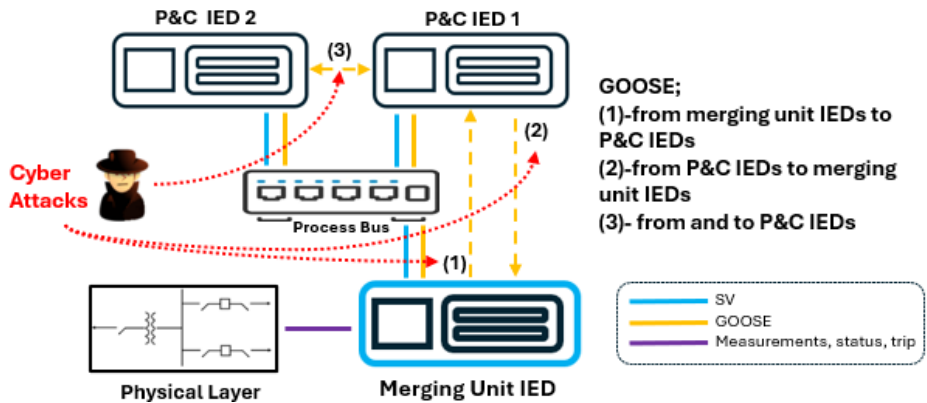

*Figure 1 – GOOSE messages in an IEC 61850-based substation.*

## 3. GOOSE CYBER ATTACK SCENARIOS

The frame structure of a GOOSE packet consists of GOOSE protocol data unit (PDU) encapsulated directly into an Ethernet frame. GOOSE PDU consists of APPID, Length, Reserved1, Reserved2 and GOOSE APDU fields. The status number (*stNum*) and sequence number (*sqNum*) in the APDU vary with the change of the states represented by the GOOSE message and for each reception of a GOOSE message, respectively. Figure 2 demonstrates the receipt of GOOSE messages at an IED subscribed to a GOOSE stream. When there's no event, GOOSE messages are in a steady state in which the messages are sent at a fixed sampling rate ($t_1$). At steady state, *stNum* is fixed and *sqNum* is incremented, e.g., in Figure 2, 0(0) that indicates *stNum* = 0 and *sqNum* = 0, is changed to 0(1) in the next packet. When there is an event reported, GOOSE sending occurs in a burst mode, in which the sampling time is reduced to its minimum ($t_0$). Gradually, the sampling time increased by doubling the previous time with each packet and reaching $t_1$. The first message sent with the reporting of the event has a new *stNum* and *sqNum* is set to zero. Based on the ways in which an adversary may interfere with the normal operation of GOOSE, several attack scenarios can be defined, and these are discussed below.

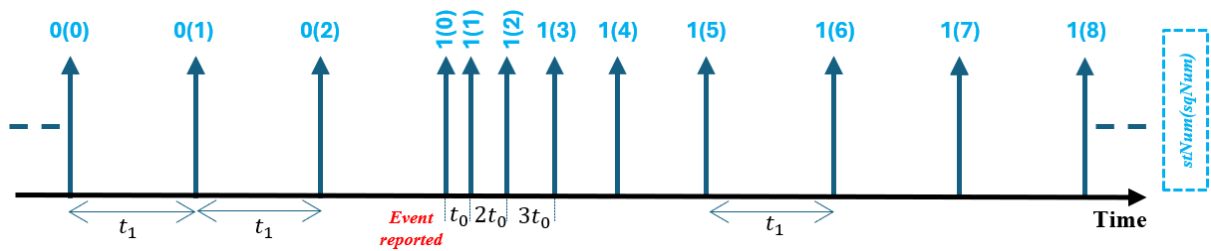

*Figure 2 – Time-series of GOOSE message reception at an IED.*

## 3.1 Replay Attack

The simplest form of GOOSE attack that an adversary can deploy is the replay attack. A GOOSE packet is captured from the substation LAN and sent back to the network after a certain time without making any changes to the original packet. As demonstrated in Figure 3, the packet labelled 1(0) captured when the event occurs is transmitted after the GOOSE transmission reaches a steady state. With this type of attack, an adversary can send a GOOSE packet with a previous trip command, when the system is in normal operation and cause a malicious breaker operation.

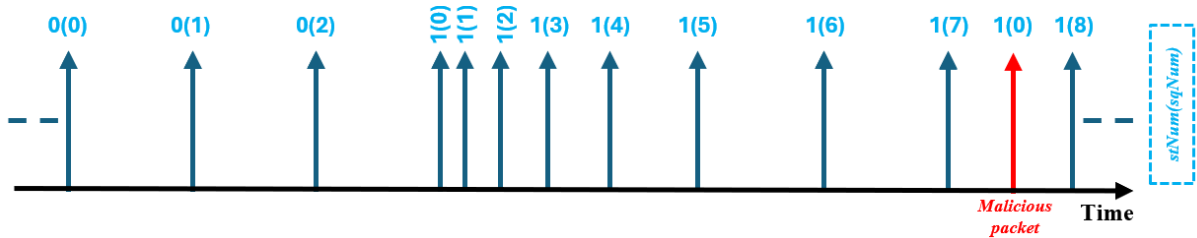

*Figure 3 – Time-series of GOOSE message reception at an IED under a replay attack.*

## 3.2 Masquerade Attack

For the implementation of a masquerade attack, an adversary should be aware of the *stNum* and *sqNum* of the most recently transmitted GOOSE packet. In this case, an adversary captures a previous GOOSE packet and updates the *stNum* and *sqNum* to see it as a legitimate packet. The adversary sends a GOOSE packet with an incremented *stNum* from the previous and *sqNum* of 0 as depicted in Figure 4. The APDU field that represents any event status is updated to 'TRUE'. Hence, when a merging unit IED receives a malicious GOOSE message that reports a protection event or a control command, it may trigger breaker operations. This attack type can evade the mechanisms deployed to detect the replay attacks.

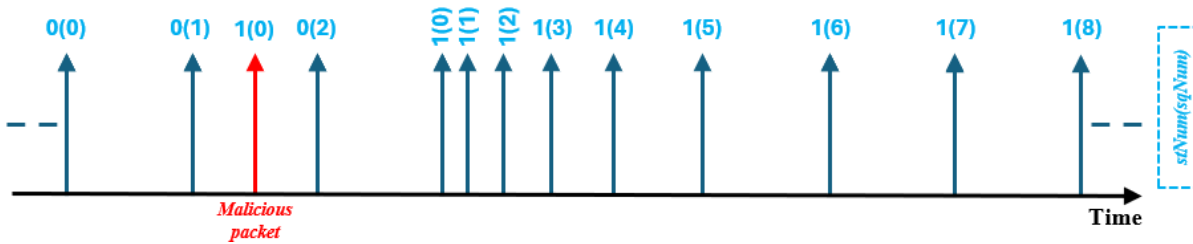

*Figure 4 – Time-series of GOOSE message reception at an IED under a masquerade attack.*

## 3.3 Flooding DoS

A flooding DoS attack on GOOSE can be launched by injecting a high volume of malicious GOOSE packets into the substation LAN, thereby overwhelming the bandwidth or the processing capacity of IEDs. In this type of attack, the adversary does not manipulate the *stNum* or *sqNum* values ; instead, it relies on saturating the communication channel so that legitimate GOOSE frames are delayed or dropped. As illustrated in Figure 5, the original GOOSE traffic is buried under a surge of malicious packets. This attack can significantly degrade the timely delivery required by IEC 61850 or even completely disrupt the delivery of legitimate packets. Flooding attacks typically inject GOOSE frames at rates far exceeding the normal steady-state retransmission interval, which is usually on the order of hundreds to thousands of packets per second. Even at approximately 1000 Hz, such an attack can overwhelm the bandwidth reserved for GOOSE messaging, since in a tightly designed network, GOOSE traffic is normally well below 100 Hz.

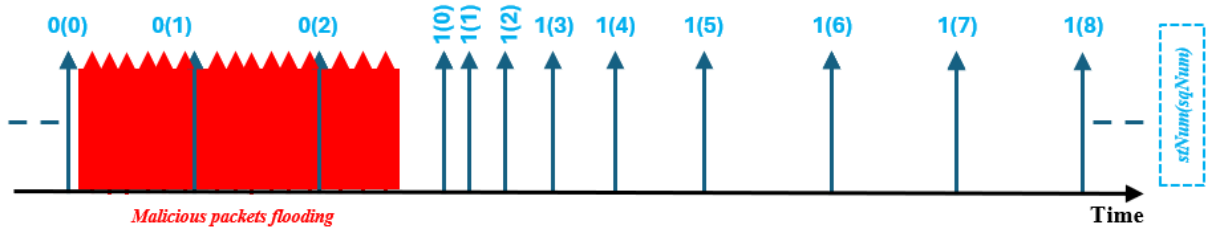

Figure 5 – *Time-series of GOOSE message reception at an IED under a flooding DoS attack.*

### 3.4 Packet Drop DoS

Unlike other GOOSE attacks that focus on capturing and injecting frames into the LAN, this attack involves direct manipulation of an Ethernet switch or the process bus to drop legitimate traffic. In a packet drop DoS attack, an adversary disrupts the forwarding path so that GOOSE packets are selectively discarded before reaching the subscribing IEDs. This can be realized by overloading switch buffers with malicious traffic, exhausting internal resources, or maliciously enforcing drop rules through a compromised configuration in a software-defined networking (SDN) controller. As a result, subscribers fail to receive timely GOOSE updates. Figure 6 illustrates a scenario in which the first few packets of an event are dropped, potentially causing a delayed breaker operation that can lead to damage to the physical system. The critical issue is that if packet dropping is extended until the end of the *timeAllowedToLive* interval of GOOSE, the communication will be declared a failure.

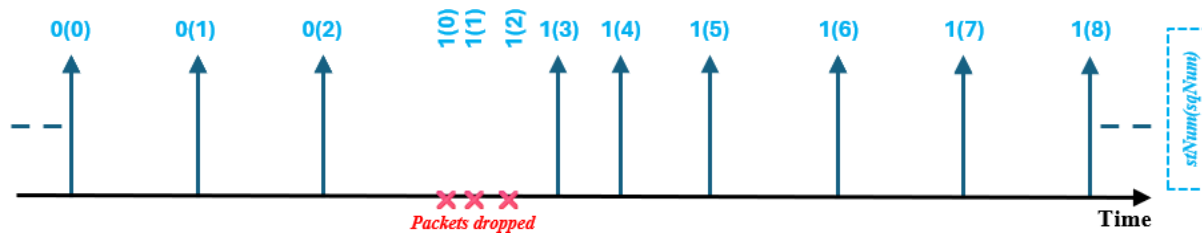

Figure 6 – *Time-series of GOOSE message reception at an IED under a packet drop DoS attack.*

## 4. REAL-TIME MITIGATION TECHNIQUES FOR GOOSE CYBER ATTACKS

### 4.1 IEC 62351 Compliant MAC verification

To address cyber security concerns in IEC 61850-based communication networks, IEC TC 57 published the IEC 62351 standard. IEC 62351-6:2020 recommends the use of message authentication code (MAC)-based algorithms that rely on pre-shared keys and offer significantly lower computational overhead [4]. When MAC-based authentication is applied to GOOSE messages, extension fields are appended to the original GOOSE packet. A secured GOOSE packet with the extension fields is illustrated in Figure 7. An initialization vector (*IV*) is appended to ensure that every MAC computation is unique, even if the payload is similar.

The key ID is the shared key among the sender and receiver IEDs. The MAC tag in the extension fields is generated by the chosen MAC algorithm.

The authentication process at the sender and receiver IEDs is illustrated in Figure 8. The sender IED generates a MAC tag by applying a MAC algorithm $f_{MAC}$ to the GOOSE PDU, using the shared key $k$ and an initialization vector ($IV$). The MAC tag is then appended to the GOOSE packet before transmission. Upon receipt, the receiver IED recalculates the MAC tag by applying $f_{MAC}$ with $k$ and the $IV$ extracted from the packet. The authenticity of the packet is verified by comparing the calculated tag with the received tag. If the packet is unauthentic, a flag is raised, and the packet is discarded. In this study, the AES-GMAC-128 algorithm, which is the primary recommendation in IEC 62351-6, is used as $f_{MAC}$.

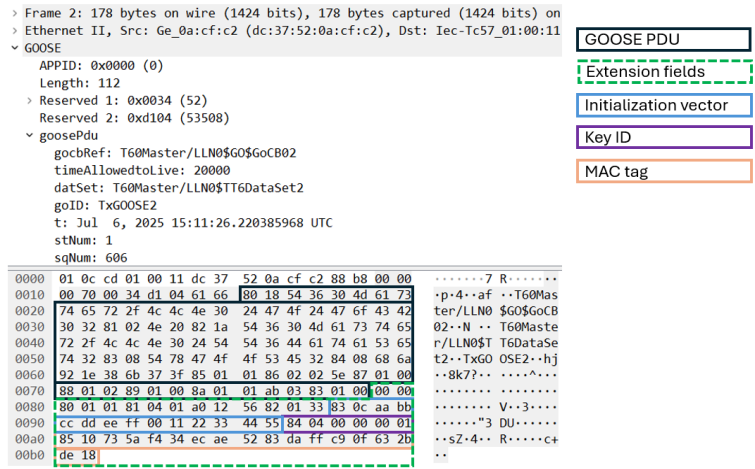

*Figure 7 – GOOSE packet with MAC verification extension fields.*

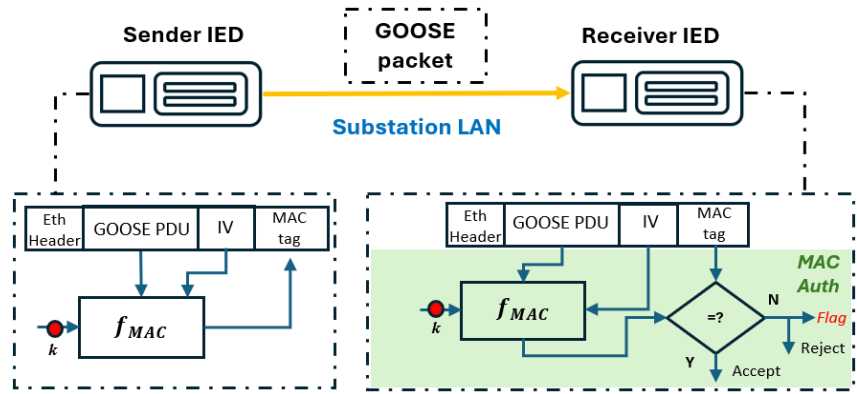

*Figure 8 – IEC 62351 compliant MAC verification procedure.*

**4.2 GOOSE Semantics Enforced Rule-Based IDS**

Malicious GOOSE traffic can be effectively detected by enforcing rule sets derived from the standardized operational behavior of GOOSE communication. These rules are primarily centered on the semantics of the *stNum* and *sqNum* fields, where their progression and data consistency are verified in accordance with the expected behavior illustrated in Figure 2. In addition, temporal validation is incorporated by utilizing a local timestamp generated upon the receipt of each GOOSE message. Since the minimum and maximum retransmission intervals are predefined, the expected number of messages within a specified observation window can be calculated and validated. Any deviation from these temporal or semantic rules is flagged as anomalous, allowing malicious packets to be identified and discarded. As the rule-based evaluation requires only lightweight parsing of packet fields without complex computations, detection can be performed in real time, ensuring minimal impact on the stringent latency requirements of IEC 61850-based communication.

As demonstrated in Figure 9, sender IED sends the original packet without any extension. At the receiver IED, *stNum* and *sqNum* fields are initially captured from the PDU of the received GOOSE packet. These fields plus a timestamp obtained at the receipt of the packet are inserted to the rule-based IDS algorithm ($g_{IDS}$) which follows the rules given in [1]. If the $g_{IDS}$ test is passed, the packet is accepted; otherwise, it is rejected and a flag is raised to indicate detection.

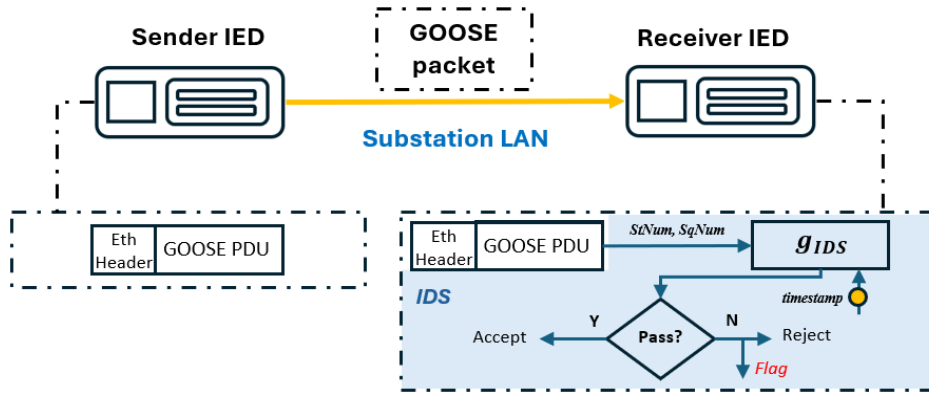

*Figure 9 – GOOSE semantics enforced rule-based IDS involved mitigation procedure.*

### 4.3 Hybrid Approach

The formulation of a hybrid process that integrates rule-based intrusion detection with MAC verification enhances the overall mitigation capability by addressing the shortcomings inherent in each individual approach. Importantly, this integration addresses the replay attack protection scheme proposed in IEC 62351-6 as an extended step for the default MAC verification process [4].

The operational flow of the hybrid approach is depicted in Figure 10. At the sender IED, the process follows the IEC 62351-defined MAC verification procedure. At the receiver IED, each incoming GOOSE message is first subjected to MAC tag verification using the function $f_{MAC}$. Upon successful authentication, the packet is further evaluated by the rule-based IDS function $g_{IDS}$. Only packets that satisfy both verification stages are accepted, while all others are discarded and flagged to indicate intrusions.

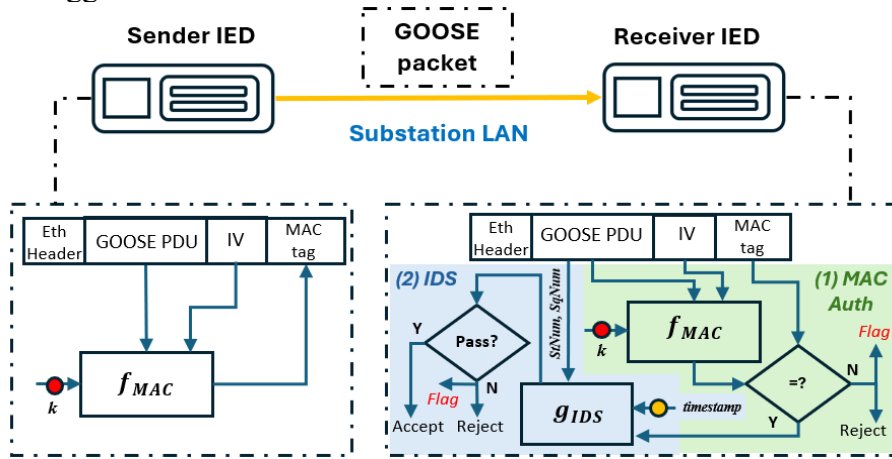

*Figure 10 – Hybrid approach's mitigation procedure.*

## 5. RESULTS AND DISCUSSION

### 5.1 Test Procedure

The CPS security testbed at Virginia Tech, as shown in Figure 11, is utilized for testing various scenarios. The physical layer of the power system is modeled using a real-time digital simulator. The MUIED that receives measurements from the real-time digital simulator via an amplifier translates the hardwired communication signals into Ethernet-based data packets and vice versa. The data sent by MUIED is received by two industry-grade P&CIEDs. At the station level, a Remote Terminal Unit (RTU) with an integrated Human-Machine Interface (HMI) is deployed. A Global Positioning System (GPS) clock is used to synchronize substation devices. To simulate cyber threats, an attacker module programmed using an open-source IEC 61850 library is implemented. A simulated IED that can perform MAC verification and intrusion detection, while replicating the functionality of a P&C IED, is included in the testbed. The P&C IED functionalities of the simulated IED are also programmed by an open-source IEC 61850 library. Within the simulated IED, a Mininet environment is used to define the substation network topology and handle GOOSE traffic [5].

The process followed in the testbed for the test cases is illustrated in Figure 12. The GOOSE messages of P&C IED 1 are sent to the process bus through a MAC generator module to append MAC tags for GOOSE packets. When the MAC verification is not tested, this element is bypassed. The MAC generator module is a computer program that utilizes the Scapy and Crypto libraries of Python to generate MAC tags. The attacker module connected to the process bus captures GOOSE messages from the process bus and publishes malicious GOOSE messages. The publisher and subscriber models of GOOSE in the IEC 61850 library are utilized for implementing the previously introduced cyber attacks, except for the packet drop DoS.

As shown in Figure 12, the GOOSE traffic received within the Mininet environment of the simulated IED is forwarded through an OpenFlow switch to a Ryu controller, where the MAC verification and intrusion detection mechanisms are implemented. Verified GOOSE packets are then delivered from the controller to the P&C IED program in the simulated IED, while malicious packets are dropped. The packet drop DoS attack is simulated from the OpenFlow switch by not allowing the attacked packets to reach the Ryu controller. The detection and mitigation of the tested cyber attacks are performed at the MAC verification and IDS module implemented inside the Ryu controller, whereas the latency is evaluated within the Mininet environment.

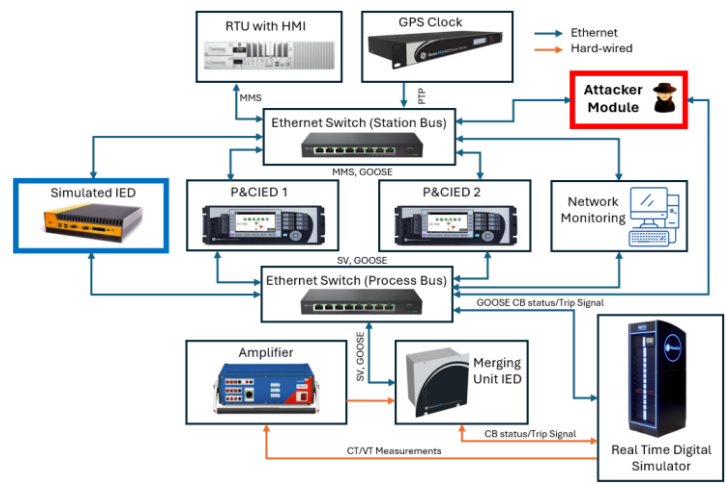

*Figure 11 – CPS security testbed.*

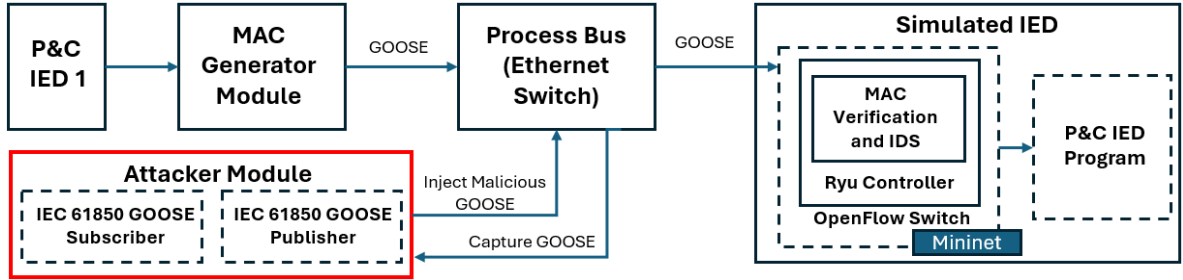

*Figure 12 – Process followed in the test cases.*

### 5.2 Test Results for Cyber Attacks

The test results demonstrated that IEC 62351-compliant MAC verification alone is effective against masquerade and flooding attacks. Figure 13 shows a capture of simulated IED interface when the MAC verification dropped the masquerade attack packet. On the other hand, a rule-based IDS that inspects packet semantics can independently detect replay and supplement mitigation. The integration of rule-based IDS with MAC verification extends the scope of MAC verification by enabling replay attack detection. Figure 14 illustrates the scenario of the hybrid approach, where the replay attack packet is detected and dropped at the IDS stage. This validates the additional *stNum* and *sqNum* checking specified in IEC 62351-6:2020 for defense against replay attacks.

For packet-drop DoS, neither the rule-based IDS, MAC verification, nor their hybrid approach provided mitigation, although the rule-based IDS was able to detect the incident. Effective mitigation requires coupling the IDS with a traffic-control mechanism, such as software-defined networking (SDN). In addition, the provision of network path redundancy through PRP/HSR protocols can serve as a potential solution.

The observations and explanations on the detection and mitigation of the tested cyber attacks using MAC verification, rule-based IDS, and the hybrid approach are presented in Table 1, Table 2, and Table 3, respectively.

*Table 1: MAC verification technique - responses to cyber attacks.*

| Attack type | Detection | Mitigation | Observation/ Explanation |
|---|---|---|---|
| Replay | Fail | Fail | The injected malicious packet is a previously captured packet with a MAC tag generated by the shared key. When it is tested for authenticity, it matches the computed tag at the receiver, as no data change in the GOOSE PDU. |
| Masquerade | Pass | Pass | The *stNUm* and *sqNum* fields in the original packet's GOOSE PDU are tampered. When $f_{MAC}$ is applied to the tampered PDU, a mismatch in the compared tags can be observed. |
| Flooding | Pass | Pass | The unauthentic malicious packets injected at 1000Hz frequency are rejected by performing the authentication for each without any congestion. |
| Packets drop | Fail | Fail | The packets received after the dropped packets are accepted as they are legitimate. However, no flag regarding dropped packets. |

```
✓ GOOSE (01:0c:cd:01:00:10) MAC Auth PASS. Forwarding packet from dc:37:52:0a:cf:c2  ⎤
✓ GOOSE (01:0c:cd:01:00:10) MAC Auth PASS. Forwarding packet from dc:37:52:0a:cf:c2  ⎥  Normal traffic
✓ GOOSE (01:0c:cd:01:00:10) MAC Auth PASS. Forwarding packet from dc:37:52:0a:cf:c2  ⎥
✓ GOOSE (01:0c:cd:01:00:10) MAC Auth PASS. Forwarding packet from dc:37:52:0a:cf:c2  ⎦
⊘ GOOSE (01:0c:cd:01:00:10) MAC Auth FAIL. Dropping packet from dc:37:52:0a:cf:c2   →  Masquerade attack packet mitigated!
✓ GOOSE (01:0c:cd:01:00:10) MAC Auth PASS. Forwarding packet from dc:37:52:0a:cf:c2
✓ GOOSE (01:0c:cd:01:00:10) MAC Auth PASS. Forwarding packet from dc:37:52:0a:cf:c2
✓ GOOSE (01:0c:cd:01:00:10) MAC Auth PASS. Forwarding packet from dc:37:52:0a:cf:c2
```

*Figure 13 – Simulated IED interface showing the masquerade attack mitigation by MAC verification.*

```
✓ GOOSE (01:0c:cd:01:00:10) Auth + IDS PASS. Forwarding packet from dc:37:52:0a:cf:c2  ⎤
✓ GOOSE (01:0c:cd:01:00:10) Auth + IDS PASS. Forwarding packet from dc:37:52:0a:cf:c2  ⎥  Normal traffic
✓ GOOSE (01:0c:cd:01:00:10) Auth + IDS PASS. Forwarding packet from dc:37:52:0a:cf:c2  ⎥
✓ GOOSE (01:0c:cd:01:00:10) Auth + IDS PASS. Forwarding packet from dc:37:52:0a:cf:c2  ⎦
⊘ GOOSE IDS (01:0c:cd:01:00:10) FAIL. Dropping packet from dc:37:52:0a:cf:c2   →  Replay attack packet mitigated!
✓ GOOSE (01:0c:cd:01:00:10) Auth + IDS PASS. Forwarding packet from dc:37:52:0a:cf:c2
✓ GOOSE (01:0c:cd:01:00:10) Auth + IDS PASS. Forwarding packet from dc:37:52:0a:cf:c2
✓ GOOSE (01:0c:cd:01:00:10) Auth + IDS PASS. Forwarding packet from dc:37:52:0a:cf:c2
```

*Figure 14 – Simulated IED interface showing the replay attack mitigation by hybrid approach.*

*Table 2: Rule-based IDS technique- responses to cyber attacks.*

| Attack type | Detection | Mitigation | Observation/ Explanation |
|---|---|---|---|
| Replay | Pass | Pass | The *stNUm* and *sqNum* fields of the injected malicious packet is not matched with order of the previously received packet. |
| Masquerade | Fail | Fail | The injected packet follows the *stNum* and *sqNum* rules and not violates any timing rules for a new event. Thus, malicious packet is accepted. |
| Flooding | Fail | Fail | Malicious packets injected at 1000Hz frequency are initially discarded by rule violations. Eventually, IDS lost synchronization with *stNum* and *sqNum* and started to discard even the legitimate packets. |
| Packets drop | Pass | Fail | The mismatch in the sequence rules of packets received after the dropped packets, and the continuation of packet loss beyond $t_1$, triggered the intrusion detection flag. However, mitigation cannot be achieved as packets' unavailability is not solved. |

*Table 3: Hybrid approach - responses to cyber attacks.*

| Attack type | Detection | Mitigation | Observation/ Explanation |
|---|---|---|---|
| Replay | Pass | Pass | Although the replayed packet passed the MAC verification stage, it is detected at $g_{IDS}$ and rejected. |
| Masquerade | Pass | Pass | The malicious packet is detected due to failing of the MAC verification. |
| Flooding | Pass | Pass | Malicious packets injected at 1000Hz are detected and discarded at the MAC verification stage before reaching the IDS stage. |
| Packet drop | Pass | Fail | The packet drop event is flagged by the IDS, although no flag from the MAC verification. But no mitigation action is taken. |

### 5.3 Latency Evaluation

The processing times of each mitigation technique were measured at the simulated IED, and the results are presented in Table 4. To obtain these measurements, GOOSE streams containing messages in both steady-state and burst modes were consistently injected.

*Table 4: Measured processing times for each mitigation technique.*

| Technique | Average processing time/ ms | Maximum processing time/ms |
|---|---|---|
| MAC verification | 0.38 | 0.64 |
| Rule-based IDS | 0.25 | 0.40 |
| Hybrid approach | 0.54 | 0.88 |

It has been demonstrated that the latency introduced by the MAC tag generation process and the associated end-to-end delays in GOOSE communication remain well below 1 ms [6]. Accordingly, the measured processing times indicate that the total additional latency from all three mitigation techniques remains under 2 ms. This comfortably satisfies the strict 3 ms delivery requirement specified for type 1A GOOSE communication.

These measurements were obtained using a computer equipped with a modest processor and 4 GB of RAM. Furthermore, no CPU-level optimizations or platform-specific tuning were applied to the cryptographic routines; the results reflect default library and operating system settings. Therefore, with advanced hardware and dedicated optimization, the processing times could be further reduced.

## 6. CONCLUSIONS

Mitigation techniques for GOOSE-based attacks in an IEC 61850-based substation have been implemented in a simulated IED and evaluated using a CPS security testbed. The results demonstrated that IEC 62351-based MAC verification can mitigate masquerade and flooding DoS attacks, while a semantics-enforced rule-based IDS can mitigate replay attacks. A hybrid approach that integrates MAC verification with rule-based IDS extends the coverage of MAC verification to include replay attacks. In addition, the testbed simulations validated that the latencies of all the implemented techniques satisfy the strict delivery requirements for GOOSE, thereby confirming their applicability for real-time cyber attack mitigation. However, none of the techniques can provide mitigation against packet drop DoS attacks, although such incidents can be detected through rule-based IDS. Effective mitigation requires coupling detection/authentication with network availability-enhancing mechanisms. Future work should therefore explore integrating these security mechanisms with redundancy protocols (e.g., PRP/HSR) or SDN-based traffic control to ensure resilience against packet drop DoS attacks and enhance overall system availability.